\documentclass[conference]{IEEEtran}
\usepackage{lineno,hyperref}
\modulolinenumbers[5]
\IEEEoverridecommandlockouts
\usepackage{amsmath,mathtools}
\usepackage{amssymb}
\usepackage{pifont}

\usepackage{fancyhdr}

\usepackage{graphicx}
\usepackage{amsmath}
\usepackage[usenames]{color}
\usepackage{subfig}
\usepackage{algpseudocode}
\usepackage{boxedminipage}
\usepackage{fancyhdr}
\usepackage{multirow}

\usepackage{array}
\usepackage{textcomp}
\usepackage{stfloats}
\usepackage{csquotes} 
\usepackage{url}
\usepackage{verbatim}
\usepackage{cite}
\usepackage[usenames]{color}
\usepackage{eqnarray}
\def\BibTeX{{\rm B\kern-.05em{\sc i\kern-.025em b}\kern-.08em
    T\kern-.1667em\lower.7ex\hbox{E}\kern-.125emX}}
\usepackage{balance}
\usepackage{url}

\usepackage{cite}

\usepackage{amsmath}
\usepackage{listings}
\usepackage{color}
\usepackage{url}
\usepackage{graphicx}  
\usepackage{float}

\ifCLASSINFOpdf
\else
\fi
\usepackage{algorithm}
\usepackage{algpseudocode}

\usepackage[top=0.7in, bottom=1.05in, left=1in, right=1in]{geometry}

\begin{document}
%

\title{Federated Edge Learning for Predictive Maintenance in 6G Small Cell Networks}

\author{
    \IEEEauthorblockN{Yusuf Emir SEZGIN\IEEEauthorrefmark{1}, Mehmet Ozdem\IEEEauthorrefmark{2}, and Tuğçe BILEN\IEEEauthorrefmark{3}}
    
    \IEEEauthorblockA{\IEEEauthorrefmark{1}Department of Computer Engineering\\}
     \IEEEauthorblockA{\IEEEauthorrefmark{3}Turk Telekom, Istanbul,Turkey }
    \IEEEauthorblockA{\IEEEauthorrefmark{2}Department of Artificial Intelligence and Data Engineering\\Faculty of Computer and Informatics\\
    Istanbul Technical University, Istanbul, Turkey\\
    Email: sezginy20@itu.edu.tr, mehmet.ozdem@turktelekom.com.tr, bilent@itu.edu.tr}
}


%

\pagestyle{fancy}
\fancyhf{}
\fancyhead[C]{\scriptsize Accepted by 5th Workshop on Integrated, Intelligent and Ubiquitous Connectivity for 6G and Beyond in IEEE International Symposium on Personal, Indoor and Mobile Radio Communications (PIMRC), ©2025 IEEE}
\renewcommand{\headrulewidth}{0pt}

\maketitle

\begin{abstract}
The rollout of 6G networks introduces unprecedented demands for autonomy, reliability, and scalability. However, the transmission of sensitive telemetry data to central servers raises concerns about privacy and bandwidth. To address this, we propose a federated edge learning framework for predictive maintenance in 6G small cell networks. The system adopts a Knowledge Defined Networking (KDN) architecture in Data, Knowledge, and Control Planes to support decentralized intelligence, telemetry-driven training, and coordinated policy enforcement. In the proposed model, each base station independently trains a failure prediction model using local telemetry metrics, including SINR, jitter, delay, and transport block size, without sharing raw data. A threshold-based multi-label encoding scheme enables the detection of concurrent fault conditions. We then conduct a comparative analysis of centralized and federated training strategies to evaluate their performance in this context. A realistic simulation environment is implemented using the ns-3 mmWave module, incorporating hybrid user placement and base station fault injection across various deployment scenarios. The learning pipeline is orchestrated via the Flower framework, and model aggregation is performed using the Federated Averaging (FedAvg) algorithm. Experimental results demonstrate that the federated model achieves performance comparable to centralized training in terms of accuracy and per-label precision, while preserving privacy and reducing communication overhead. 
\end{abstract}

\begin{IEEEkeywords}
Federated Learning, Predictive Maintenance, 6G Networks, Small Cell, Edge Intelligence.
\end{IEEEkeywords}

\section{Introduction}
The transition to sixth-generation (6G) communication networks introduces new opportunities for high-speed, ultra-reliable, and intelligent wireless communication. With increasing demands on latency, connectivity, and autonomy, the underlying infrastructure, small cell base stations (gNBs), must operate with minimal downtime and be capable of self-maintenance. 6G is expected to enable a fully connected system by integrating cloud, sensors, devices, vehicles, and robotic agents into a pervasive communication fabric, surpassing the capabilities of 5G in terms of data rate, latency, and scalability \cite{BILEN2020101133}. In this context, predictive maintenance becomes a crucial enabler for service continuity and cost-effective network management.

Traditional approaches to predictive maintenance often rely on centralized machine learning pipelines, where telemetry data from base stations is transmitted to a central server for training \cite{10078095}. However, this method poses significant limitations in terms of data privacy, network bandwidth consumption, and scalability due to the dense and distributed 6G architecture. There is a pressing need for decentralized learning systems that can predict hardware failures locally while enabling global model coordination.

Based on these challenges, in this paper, we propose a federated edge learning approach for predicting and prioritizing hardware failures in 6G small cell networks. In this approach, each base station independently trains a failure detection model on its own telemetry data and contributes to the training of a global model without sharing raw data. Specifically, the main contributions of this paper are summarized as follows:
\begin{itemize}
    \item We propose a federated learning-based predictive maintenance framework tailored for 6G base stations. Each base station trains a local multi-label classifier using real-time telemetry data (SINR, jitter, delay, and tbSize) and contributes to a global anomaly detection model while preserving data privacy. 
    \item We design a threshold-based binarization method to label failure conditions across multiple KPIs. This enables robust multi-label classification in complex and overlapping fault scenarios.
    \item We simulate realistic 6G network behavior using ns-3 with mmWave modules and extract per-user telemetry logs via custom packet tracing to generate training data for both centralized and federated models.
\end{itemize}

The paper is organized as follows: Section~\ref{sec:related} reviews related work, Section~\ref{sec:arch} outlines the system structure, Section~\ref{sec:approach} describes the proposed approach, Section~\ref{sec:evaluation} presents results, and Section~\ref{sec:conclusion} concludes.

\section{Related Works}
\label{sec:related}
Literature includes various approaches to address the challenges of reliability and intelligence in next-generation wireless systems. One significant challenge in federated learning (FL) is managing performance degradation caused by non-independent and identically distributed (non-IID) data across clients. \cite{zhao2018federated} shows that such distributional shifts can significantly reduce global model accuracy, and proposes the exchange of small global data subsets to improve convergence. In the context of predictive maintenance, where data collected by edge nodes (e.g., base stations) varies due to environmental and traffic conditions, these methods become highly applicable. Building on this, \cite{li2021nonIID} presents an experimental framework for evaluating federated learning under realistic data silos. This work identifies strategies to reduce the adverse impact of heterogeneity through local training optimizations.

Several studies emphasize the need for distributed intelligence in edge networks to meet the low-latency and autonomy goals of 6G. \cite{taik2020feel} investigates federated edge learning (FEEL) as a paradigm for enabling collaborative intelligence in highly dynamic, resource-constrained settings. The authors propose client selection and model aggregation techniques tailored for delay-sensitive environments. Similarly, \cite{zhang2024foundation} proposes a distributed foundation model architecture for multi-modal learning in wireless networks. This work aims to maintain generalization across spatially distributed and heterogeneous edge inputs. These efforts align with 6G design principles, where predictive maintenance must operate without centralized coordination.

There is also increasing attention to scaling FL systems to support more complex models such as large language models (LLMs). \cite{piccialli2024llm} analyzes the integration of FL with edge learning for LLMs. This work highlights the balance between performance, energy consumption, and communication overhead. Their hybrid architecture leverages edge resources for data processing while coordinating global model updates across the network. These findings suggest that predictive fault detection tasks can benefit from applying LLMs when combined with efficient edge deployment strategies.

Foundational research also continues to shape the evolution of FL. \cite{mcmahan2017fedavg} introduces the Federated Averaging algorithm, which has become a cornerstone of many FL systems due to its efficiency in decentralized gradient updates. Meanwhile, \cite{yang2019federatedml} provides a broad overview of FL concepts and architectures, including vertical and horizontal FL setups, privacy preservation, and application-specific tuning. Complementing these, \cite{kairouz2021advances} reviews recent advances and outlines critical gaps in the literature, such as robustness, fairness, and personalization—areas especially relevant to predictive maintenance scenarios where each node may have unique fault profiles. On the deployment side, \cite{bonawitz2019scale} describes an industry-scale implementation of FL in mobile systems, offering practical insights into compression, communication, and security that can inform real-world 6G applications.
\\
While prior works introduce foundational methods for handling non-IID data, edge resource limitations, or large-scale deployments, they often fall short of integrating these approaches within realistic 6G network simulations. In contrast, our work uniquely bridges this gap by coupling federated edge learning with a detailed mmWave-based simulation environment and multi-label failure labeling. Unlike studies focused solely on algorithmic improvements or infrastructure design, our system is validated in an end-to-end manner (from data generation to federated training) under varying user densities and failure conditions. This holistic evaluation demonstrates the feasibility and scalability of privacy-preserving predictive maintenance in 6G base stations, making our contribution both practical and immediately applicable to next-generation network scenarios.

\section{The Proposed System Architecture}
\label{sec:arch}
To address the challenges identified in the previous section, we propose a predictive maintenance framework for 6G small cell networks based on a Knowledge Defined Network (KDN) architecture, as illustrated in Fig.~\ref{fig:system_structure}. This architecture consists of three main layers: the \textit{Data Plane}, the \textit{Knowledge Plane}, and the \textit{Control Plane}. The details of these planes could be explained as follows:

\begin{figure}[ht]
  \centering
  \includegraphics[width=0.75\linewidth]{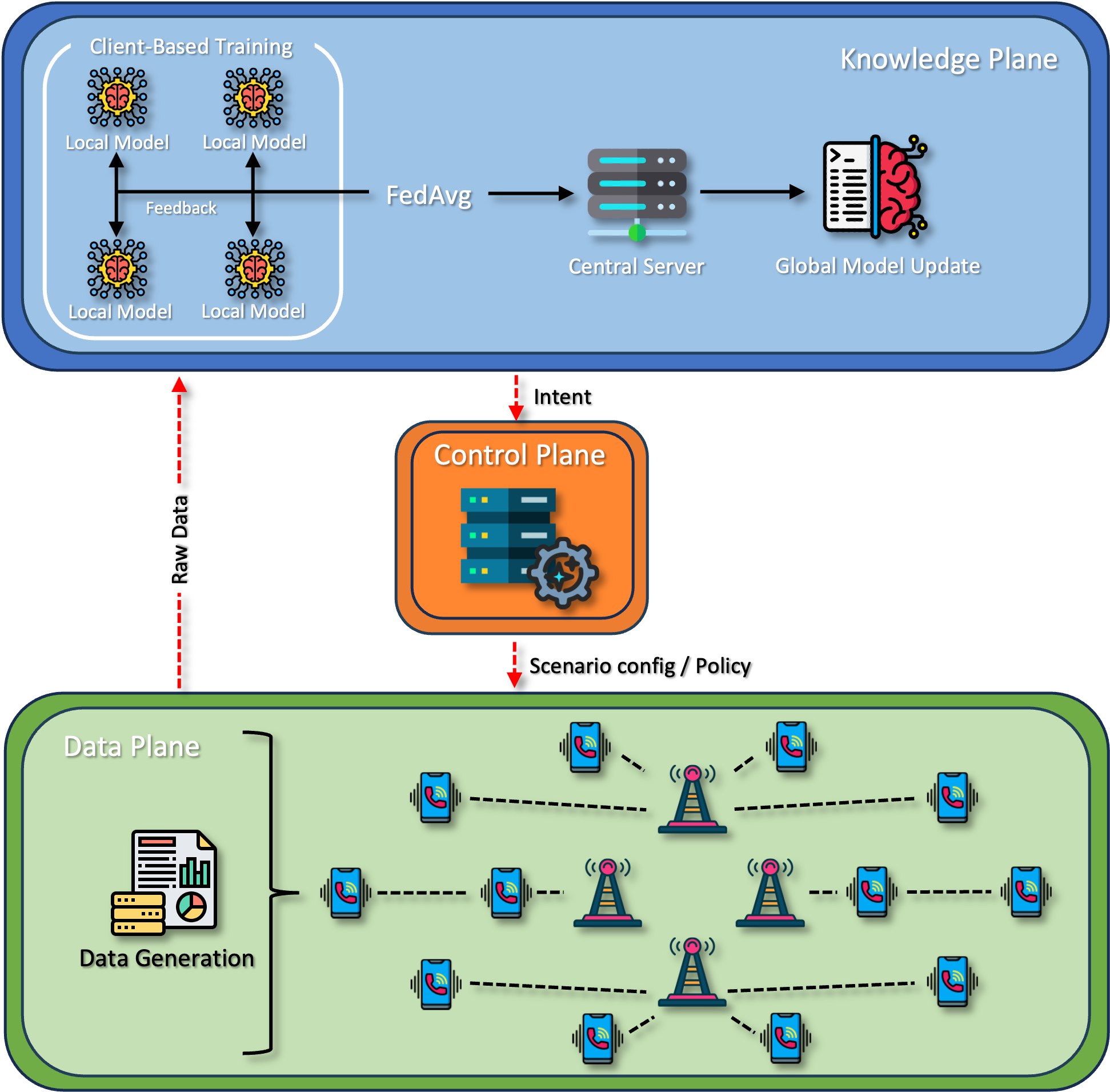}
  \caption{KDN-Based Proposed System Architecture}
  \label{fig:system_structure}
\end{figure}

\begin{itemize}
    \item Data Plane: The data plane represents the operational environment in which user devices communicate with distributed base stations. It simulates wireless interactions under variable spatial, temporal, and load conditions. The data plane is responsible for collecting telemetry that reflects network behavior, including communication events, transmission quality, and service conditions \cite{10167962}. As shown in Fig. \ref{fig:system_structure}, the control plane interfaces with the data plane to configure user deployment, placement strategies, and node behavior under normal and fault-induced conditions. The resulting telemetry is preprocessed and exported for learning within the knowledge plane.
    \item Knowledge Plane: The knowledge plane handles model training and predictive reasoning. Each base station hosts a local intelligence module that operates on its own telemetry data. These models are coordinated via a federated learning protocol, where model parameters are shared with a central server for aggregation. As shown in Fig. \ref{fig:system_structure}, the aggregated global model is redistributed to all participating nodes, enabling continual improvement over multiple communication rounds. The knowledge plane may also include mechanisms for feedback generation (e.g., predicting failure intent), which is passed to the control plane for downstream policy adaptation.
    \item Control Plane: The control plane serves as the system coordinator. It defines experimental scenarios, manages simulation parameters, and orchestrates participation in learning rounds. This includes managing client selection, triggering failure events, and scheduling model updates \cite{9927253}. Crucially, the control plane also receives predictions and alerts from the knowledge plane in the form of intent. These insights inform decisions such as retraining frequency, fault prioritization, or simulation reconfiguration. This closed feedback loop ensures that the overall system adapts dynamically to changing network behavior while maintaining scalability and data privacy.
\end{itemize}

Therefore, data flows upward from the data plane into the knowledge plane, where it supports the training of distributed models. Policy directives and training intent flow downward through the control plane to adjust simulation behavior and node participation. This layered architecture supports privacy-preserving telemetry processing, scalable model training, and dynamic coordination between components. Each plane fulfills a distinct functional role while interacting through clearly defined interfaces. Accordingly, we can satisfy the requirements of 6G networks.

\section{The Proposed Approach}
\label{sec:approach}

This section presents our predictive maintenance framework developed for 6G small cell networks using federated edge learning (FEEL). The approach is structured around four key contributions: (i) Decentralized Learning Engine for 6G gNBs, (ii) Multi-KPI Fault Encoding Engine with Adaptive Thresholding, (iii) KPI-Aware Telemetry Structuring Engine, and (iv) Dual-Pipeline Learning Engine. The details of these parts will be explained in the upcoming parts. 

\subsection{Decentralized Learning Engine for 6G gNBs}
Our system treats each 6G base station as a local intelligent node, capable of independently training a failure prediction model using its own telemetry data. The telemetry includes four key performance indicators (KPIs): SINR, jitter, delay, and transport block size (tbSize). These metrics are extracted from real-time communication logs between the base stations and user equipment (UE).

Each base station segments its telemetry into fixed-length time windows and computes statistical features (e.g., moving averages and rate of change) before passing them into a compact neural network classifier. This model is trained locally using binary cross-entropy loss in a multi-label setup.

At the end of each local training phase, the base stations transmit only their model updates to a central server, without revealing any raw telemetry. The server then aggregates these updates using the Federated Averaging (FedAvg) algorithm. It is a widely adopted method for decentralized optimization in federated learning settings. The global model is updated by computing a weighted average of the client updates, and the resulting model is broadcast back to all participants. This process repeats over multiple communication rounds until convergence is achieved.

We implement this training workflow using the Flower federated learning framework, which provides a flexible and modular infrastructure for simulating distributed clients, managing communication rounds, and coordinating model aggregation.

This design ensures both scalability and privacy. No raw data ever leaves the base station, thereby satisfying the data protection constraints of the 6G network while significantly reducing uplink communication costs. Since the learning model operates in a supervised, multi-label setup, a key requirement is the conversion of raw telemetry metrics into binary fault indicators. This process is described next.

\subsection{Multi-KPI Fault Encoding Engine with Adaptive Thresholding}
To enable multi-label classification, a structured labeling strategy is applied to the telemetry data, mapping continuous measurements to binary fault indicators. Each telemetry window may exhibit one or more failure signatures, determined by threshold-based rules applied to selected KPIs.

As explained above, the telemetry includes four key KPIs: Signal-to-Interference-plus-Noise Ratio (SINR), jitter, packet delay, and transport block size (tbSize). Thresholds for each KPI are empirically defined based on domain expertise, historical trends, and system-level performance expectations. For example, persistently low SINR or excessively high jitter within a window indicates signal degradation, while prolonged delay or unusually small tbSize may suggest scheduling inefficiencies or transmission constraints.

If any KPI exceeds (or falls below) its respective threshold, a binary label is assigned to indicate the presence of that specific degradation type. Since these conditions may co-occur, the resulting label vector captures the multi-label nature of real-world fault patterns in 6G networks. The labeling logic is modular and adaptable, allowing for dynamic adjustment of threshold values or the inclusion of additional KPIs depending on deployment-specific requirements. This flexibility ensures that the labeling strategy remains robust across diverse operating conditions and supports accurate learning in predictive maintenance pipelines. To enable effective application of the labeling strategy described above, it is essential to structure telemetry collection in a way that captures realistic user behaviors and fault dynamics. The following section details how the telemetry data is organized and enriched with fault context.

\subsection{KPI-Aware Telemetry Structuring Engine}
To support realistic and fault-aware learning in next-generation wireless systems, we develop a network topology and data collection methodology that captures user-infrastructure interactions under variable link conditions and failure scenarios. The system models a dynamic multi-cell architecture where UE connects to distributed gNBs operating in mmWave spectrum. Here, a hybrid user placement strategy is adopted: half of the users are located near their serving gNBs to represent strong-link conditions. At the same time, the rest are positioned at the coverage edge to emulate degraded connectivity. This spatial heterogeneity introduces diversity in signal quality and traffic behavior, which is critical for training robust predictive models.

\begin{algorithm}[ht]
\caption{Topology-Aware Telemetry Collection and Fault Modeling Procedure}
\label{alg:topology}
\small
\begin{algorithmic}[1]
\State Set parameters: number of base stations $B$, users $U$, total duration $T$
\State Define user placement ranges: \textit{near\_range}, \textit{far\_range}
\State Define fault injection time $t_f$ and number of gNBs to deactivate
\State Deploy $B$ gNBs in a circular or grid layout
\For{each user $u_i$ in $U$}
    \If{$i < U/2$}
        \State Place $u_i$ within \textit{near\_range} of assigned gNB
    \Else
        \State Place $u_i$ within \textit{far\_range} of assigned gNB
    \EndIf
\EndFor
\State Initialize network protocols and link configurations
\State Assign IP addresses and configure routing tables
\State Initiate traffic flows and configure data sinks
\State Enable telemetry tracing for KPIs: SINR, jitter, delay, tbSize
\If{fault injection is enabled}
    \State At time $t_f$, deactivate selected gNBs to simulate faults
\EndIf
\State Execute runtime operations until time $T$ is reached
\State Export collected telemetry in a structured format
\end{algorithmic}
\end{algorithm}

Throughout the communication process, telemetry data is recorded for each UE–gNB interaction. As explained above, the monitored KPIs include Signal-to-Interference-plus-Noise Ratio (SINR), jitter, packet delay, and transport block size (tbSize). These metrics are collected in a structured format using tracing mechanisms embedded within the communication stack. Additionally, to simulate realistic network degradation, the methodology incorporates fault injection by selectively deactivating one or more gNBs during runtime. This enables the observation of propagation effects on performance metrics and facilitates the development of predictive maintenance and fault detection mechanisms. The complete telemetry collection and fault modeling process is outlined in Algorithm~\ref{alg:topology}. This procedure governs the deployment of base stations, the positioning of users, the tracing of telemetry, and the injection of faults within the 6G environment.

The structured and fault-enriched telemetry generated in this stage provides the basis for training predictive models. The next section explains how this data is consumed in two alternative training pipelines (centralized and federated) enabling robust learning under different deployment constraints.

\subsection{Dual-Pipeline Learning Engine}
The framework supports both centralized and federated learning pipelines. In the centralized setting, telemetry from all base stations is aggregated into a unified dataset and used to train a single model. In the federated setting, each base station performs local training on its own telemetry and contributes model updates to the global model without sharing raw data.

Both pipelines use a shared model architecture configured for multi-label classification. Training is performed using standard gradient-based optimization algorithms over a fixed number of epochs. The federated pipeline proceeds through multiple communication rounds, during which model updates are exchanged and averaged.

This dual-path training and evaluation strategy enables a fair and structured comparison between centralized and decentralized approaches, while ensuring that privacy-preserving learning remains viable for 6G network environments.

\section{Performance Evaluation}
\label{sec:evaluation}

\subsection{Simulation Details}
To evaluate the proposed predictive maintenance framework, we perform extensive simulations across various small cell deployment scenarios. The evaluation focuses on three key aspects: the failure labeling methodology, the telemetry data generation process, and the comparison between centralized and federated learning (FL) strategies. First, we introduce a threshold-based labeling approach to convert raw telemetry metrics into binary fault indicators. Each telemetry window is assessed independently and assigned a multi-label vector, where each binary indicator corresponds to a specific type of performance degradation. This labeling process enables scalable and interpretable training for both centralized and FL settings. Next, we describe the simulation setup used to generate realistic telemetry data under diverse network conditions. This environment captures different failure scenarios to test the framework's robustness and adaptability. Finally, we evaluate the performance of centralized and federated learning strategies in terms of scalability, privacy preservation, and predictive accuracy. This comparative analysis provides insights into the trade-offs between local training and centralized aggregation in real-world deployments.

Based on these, Table~\ref{table:thresholds} lists the threshold values applied to four key performance indicators (KPIs): Signal-to-Interference-plus-Noise Ratio (SINR), jitter, delay, and transport block size (tbSize). These thresholds are derived from historical performance data and system-level benchmarks to reflect abnormal network behavior. Each telemetry window is independently evaluated and assigned a multi-label vector, capturing the simultaneous presence of different degradation conditions. This lightweight labeling strategy enables the model to learn from complex and overlapping fault patterns, providing a scalable and effective basis for predictive maintenance in 6G networks.

\begin{table}[ht]
\centering
\caption{Thresholds for Label Generation}
\label{table:thresholds}
\renewcommand{\arraystretch}{1}
\scriptsize
\begin{tabular}{|c|c|c|}
\hline
\textbf{Metric} & \textbf{Threshold} & \textbf{Failure Condition} \\
\hline
SINR & $<$ 20 dB & Low signal quality \\
Jitter & $>$ 0.1 ms & Irregular packet arrival \\
Delay & $>$ 0.8 ms & High latency \\
tbSize & $<$ 500 bytes & Inefficient transmission \\
\hline
\end{tabular}
\end{table}

Also, to generate realistic training and evaluation data, we designed a custom simulation scenario in ns-3 using the mmWave module. The simulated network comprises 5, 10, or 50 base stations and 10–100 user devices, randomly distributed over a circular area with a diameter of 200 meters. To model both typical and edge-case conditions, half of the users are positioned near their associated base stations (~10 meters), while the rest are placed at longer distances (~300 meters). This hybrid spatial distribution enables us to analyze how proximity impacts telemetry quality and model robustness under varying signal conditions.

Following the initial simulation setup, we introduce controlled fault scenarios by triggering base station failures. At 0.5 seconds into each run, one to ten gNBs are forcefully disconnected by disabling their receive callback function, simulating abrupt node-level outages. This allows us to evaluate the responsiveness of the proposed models under sudden network disruptions. Here, telemetry data is collected through packet tracing callbacks in ns-3, recording SINR, delay, jitter, and transport block size (tbSize) for each user over time. The collected data is exported in CSV format, labeled using the predefined KPI thresholds (Table~\ref{table:thresholds}), and segmented into fixed-length telemetry windows. For both centralized and federated learning settings, each scenario’s dataset is randomly split into 80\% for training and 20\% for evaluation, ensuring a consistent basis for comparative analysis.

\begin{figure*}[h]	
	\centering		
        \subfloat[]{%
		\includegraphics[width=0.33\textwidth]{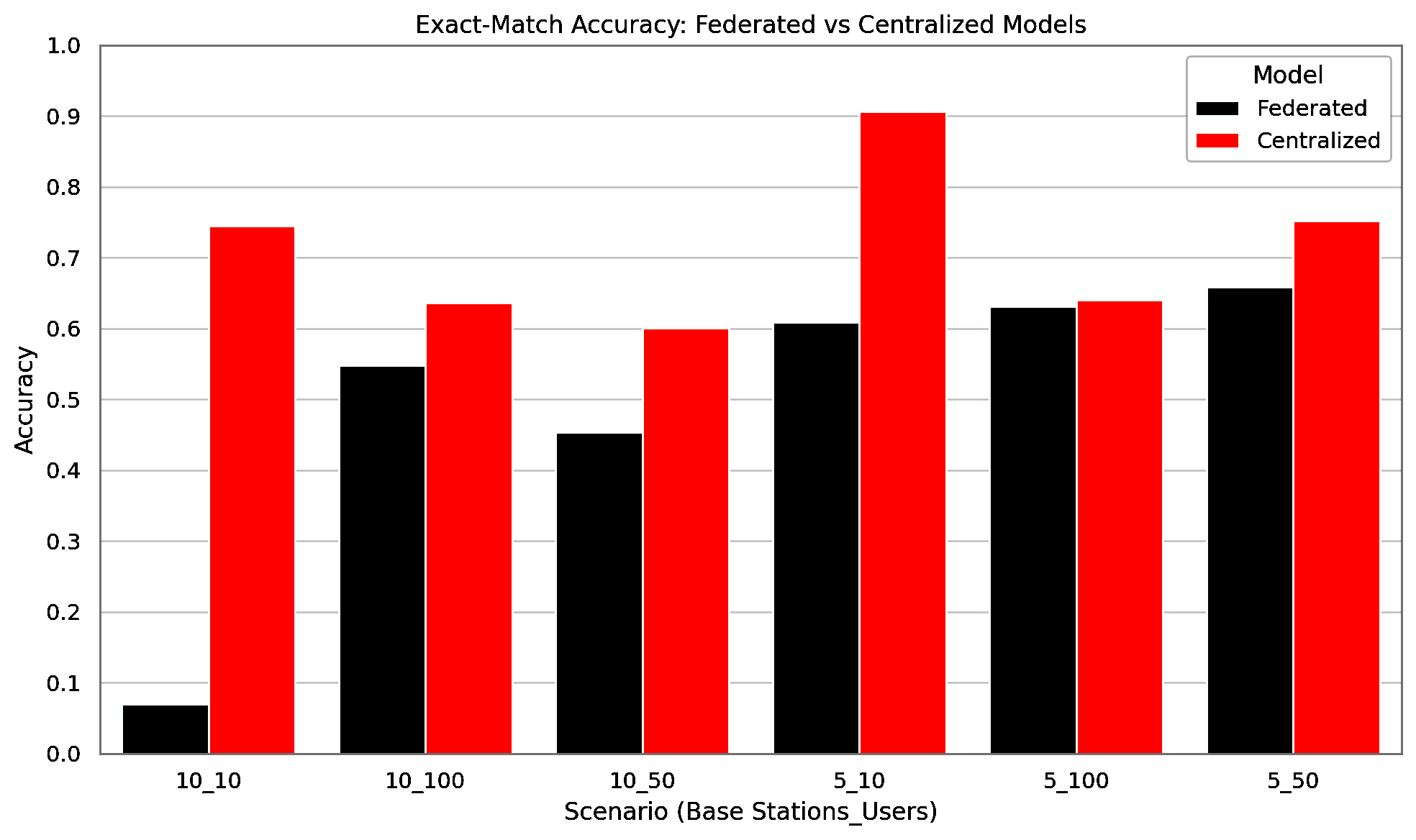}%
		\label{fig:exactmatch}%
	} 
     \subfloat[]{%
		\includegraphics[width=0.33\textwidth]{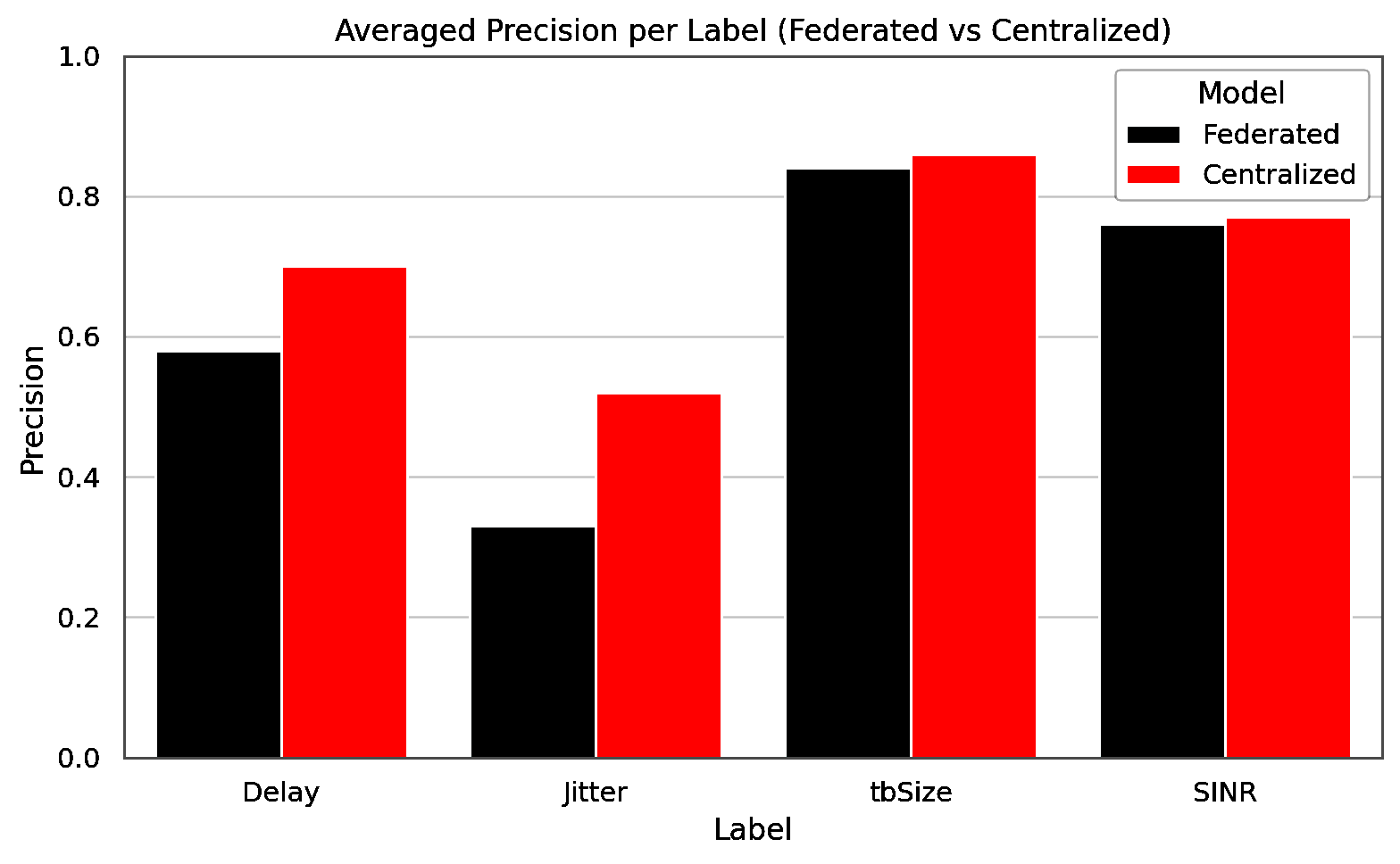}%

		\label{fig:precision}%
	} %
         \subfloat[]{%
		\includegraphics[width=0.33\textwidth]{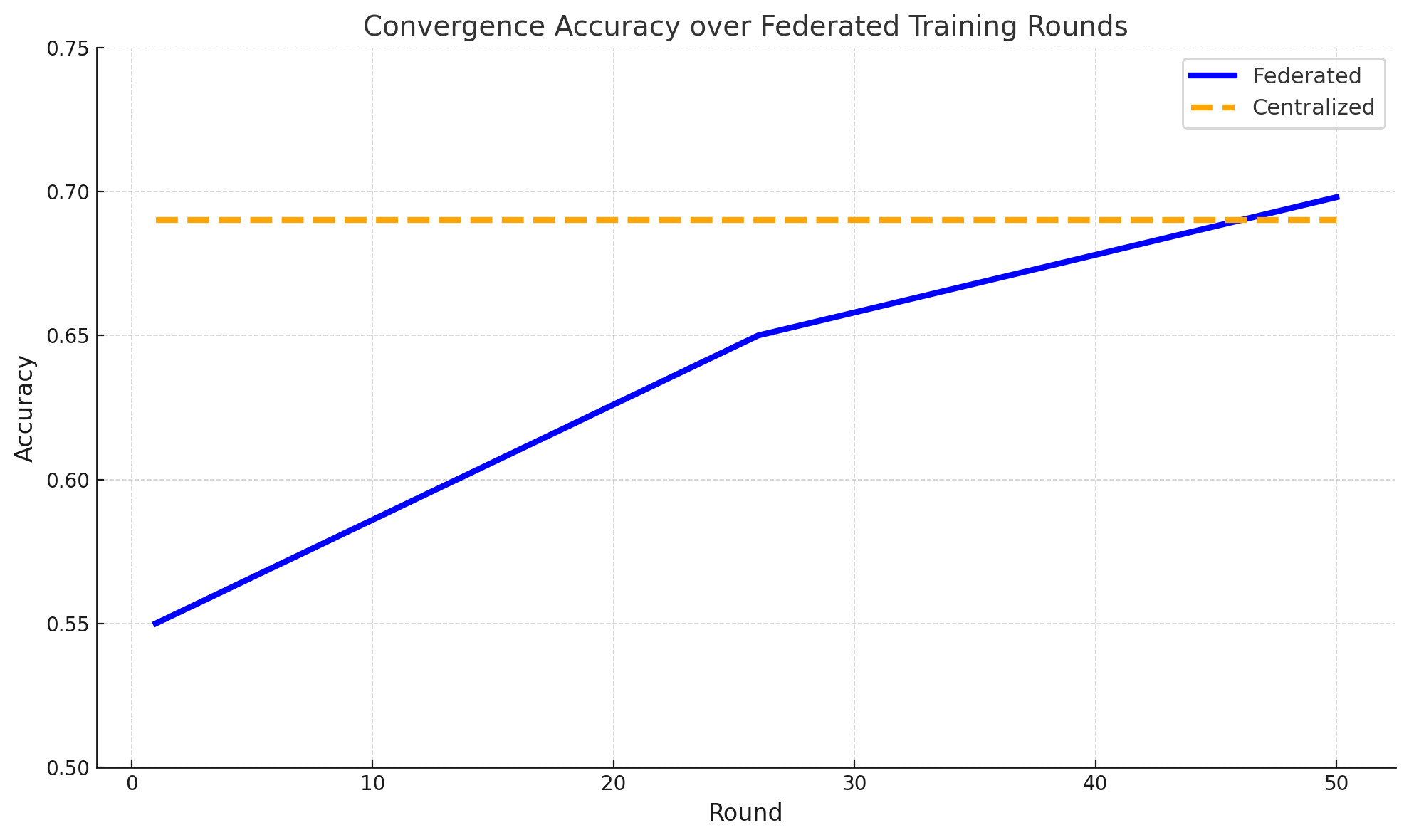}%
		\label{fig:convergence}%
	} %
	\caption{Aggregated Efficiency Index and Accuracy Analysis}
	\label{r6}
\end{figure*}

\subsection{Results and Analysis}
To benchmark our federated learning framework, we implemented a centralized training pipeline using the same neural network architecture and telemetry dataset. In the centralized setting, all telemetry data is aggregated into a single dataset for global training. Both approaches employ an identical feedforward neural network consisting of two hidden layers (64 and 32 neurons), ReLU activations, and a sigmoid output layer for multi-label classification. Models are trained using the binary cross-entropy loss function and the Adam optimizer. In the centralized setup, training is conducted for 10 full epochs on the aggregated dataset. For federated learning, we adopt the Flower framework to simulate 50 communication rounds. In each round, the current global model is distributed to participating base stations (clients), which perform 10 local training epochs on their private datasets. The locally updated weights are then sent back and aggregated by the server using the FedAvg algorithm. This process iterates until convergence. Then, we evaluate the performance of the proposed system through three key analyses: exact-match accuracy, per-label classification performance, and training convergence behavior.

\begin{itemize}
    \item \textbf{Exact-Match Accuracy:} Figure~\ref{fig:exactmatch} presents the exact-match accuracy across six simulation scenarios with varying numbers of base stations and user devices. As expected, centralized models consistently achieve higher exact-match performance due to having access to the full aggregated dataset. However, the federated models demonstrate highly competitive accuracy, particularly in denser user deployments (e.g., S3 and S6), where the volume of local training data improves model generalization. This trend illustrates that federated learning can approach centralized performance even under heterogeneous client conditions.

    \item \textbf{Per-Label Precision:} Figure~\ref{fig:precision} shows the average precision achieved per failure label. Both models achieve high precision on easier-to-detect conditions such as \textit{tbSize} and \textit{SINR}, with minimal performance differences between centralized and federated approaches. In contrast, precision drops for more complex or noisy conditions like \textit{jitter} and \textit{delay}, where centralized models retain a slight advantage. Nonetheless, federated models maintain reliable classification performance, with average precision exceeding 0.75 for most fault types, validating their robustness under distributed training constraints.
    
    \item \textbf{Training Convergence:} Figure~\ref{fig:convergence} illustrates the convergence of federated training accuracy over 50 communication rounds. The federated model shows consistent improvement and eventually stabilizes near the centralized model’s accuracy, indicating strong convergence and learning stability. This behavior reinforces the viability of decentralized learning in environments where training must occur over multiple communication cycles.
\end{itemize}

A major strength of the proposed framework is its privacy-preserving design. In contrast to centralized approaches that require raw telemetry data aggregation, the federated model operates entirely on-device and shares only model parameters. Experimental results demonstrate that this decentralized learning strategy achieves comparable predictive performance to centralized training. As a result, federated edge learning emerges as a practical solution for 6G networks, especially in scenarios constrained by data privacy regulations, communication overhead, or limited backhaul infrastructure.

\section{Conclusion}
\label{sec:conclusion}
This paper introduced a federated edge learning model for predictive maintenance in 6G small cell networks. In this model, we leverage a KDN architecture to enable decentralized training on local telemetry while preserving data privacy and reducing communication overhead. Also, we implemented a multi-label fault labeling strategy and evaluated the system across realistic simulation scenarios using ns-3. Results show that the federated approach achieves predictive performance comparable to centralized learning, while offering significant advantages in scalability and privacy.

\bibliographystyle{unsrt}
\bibliography{main}

\begin{thebibliography}{10}

\bibitem{BILEN2020101133}
Tuğçe Bilen and Berk Canberk.
\newblock Overcoming 5g ultra-density with game theory: Alpha-beta pruning
  aided conflict detection.
\newblock {\em Pervasive and Mobile Computing}, 63:101133, 2020.

\bibitem{10078095}
Tuğçe Bilen, Berk Canberk, and Trung~Q. Duong.
\newblock Digital twin evolution for hard-to-follow aeronautical ad-hoc
  networks in beyond 5g.
\newblock {\em IEEE Communications Standards Magazine}, 7(1):4--12, 2023.

\bibitem{zhao2018federated}
Yue Zhao, Meng Li, Liangzhen Lai, Naveen Suda, Damon Civin, and Vikas Chandra.
\newblock Federated learning with non-iid data.
\newblock {\em arXiv preprint arXiv:1806.00582}, 2018.

\bibitem{li2021nonIID}
Qiang Li, Yansong Diao, Qiang Chen, and Bingsheng He.
\newblock Federated learning on non-iid data silos: An experimental study.
\newblock {\em arXiv preprint arXiv:2102.02079}, 2021.

\bibitem{taik2020feel}
Ahmed Taïk and Salima Cherkaoui.
\newblock Federated edge learning: Design issues and challenges.
\newblock {\em arXiv preprint arXiv:2009.00081}, 2020.

\bibitem{zhang2024foundation}
Zonglin Zhang, Yonggang Wen, Hongyang Lyu, et~al.
\newblock Distributed foundation models for multi-modal learning in 6g wireless
  networks.
\newblock {\em IEEE Wireless Communications}, 2024.

\bibitem{piccialli2024llm}
Francesco Piccialli, Daniele Chiaro, Pengfei Qi, et~al.
\newblock Federated and edge learning for large language models.
\newblock {\em Information Fusion}, 117:102840, 2024.

\bibitem{mcmahan2017fedavg}
H~Brendan McMahan, Eider Moore, Daniel Ramage, et~al.
\newblock Communication-efficient learning of deep networks from decentralized
  data.
\newblock {\em arXiv preprint arXiv:1602.05629}, 2017.

\bibitem{yang2019federatedml}
Qiang Yang, Yang Liu, Tianjian Chen, and Yongxin Tong.
\newblock Federated machine learning: Concept and applications.
\newblock {\em ACM Transactions on Intelligent Systems and Technology (TIST)},
  10(2):12, 2019.

\bibitem{kairouz2021advances}
Peter Kairouz, H~Brendan McMahan, et~al.
\newblock Advances and open problems in federated learning.
\newblock {\em Foundations and Trends in Machine Learning}, 14(1--2):1--210,
  2021.

\bibitem{bonawitz2019scale}
Keith Bonawitz, Hubert Eichner, Walter Grieskamp, et~al.
\newblock Towards federated learning at scale: System design.
\newblock In {\em Proceedings of the 2nd SysML Conference}, 2019.

\bibitem{10167962}
Lal~Verda Cakir, Tuğçe Bilen, Mehmet Özdem, and Berk Canberk.
\newblock Digital twin middleware for smart farm iot networks.
\newblock In {\em 2023 International Balkan Conference on Communications and
  Networking (BalkanCom)}, pages 1--5, 2023.

\bibitem{9927253}
Tuğçe Bilen, Elif Ak, Bahadır Bal, and Berk Canberk.
\newblock A proof of concept on digital twin-controlled wifi core network
  selection for in-flight connectivity.
\newblock {\em IEEE Communications Standards Magazine}, 6(3):60--68, 2022.

\end{thebibliography}

\end{document}